\documentclass{amsart}

\usepackage{epsfig}
\usepackage{latexsym}
\usepackage{amsmath}
\usepackage{natbib}

\newtheorem{theorem}{Theorem}[section]
\newtheorem{proposition}[theorem]{Proposition}

\newcommand{\p}{\mathcal P}

\newcommand{\T}{\mathcal T}

\newcommand{\PP}{{\mathbb P}}

\newcommand{\old}[1]{{}}

\title{Maximum likelihood supertrees}
\author{Mike Steel and Allen Rodrigo}

\thanks{We thank the Allan Wilson Centre for Molecular Ecology and Evolution for supporting this work. AR began this project while he was working with Olivier Gascuel at the Laboratoire d'Informatique, de Robotique et de Microelectronique de Montpellier.}

\address{MS: Allan Wilson Centre for Molecular Ecology and Evolution, Department of Mathematics and
  Statistics, University of Canterbury, Christchurch, New Zealand}

\email{m.steel@math.canterbury.ac.nz}

\address{AR: Bioinformatics Institute, University of Auckland, New Zealand,
and Laboratoire d'Informatique, de Robotique et de Microelectronique
de Montpellier, France.}

\email{a.rodrigo@auckland.ac.nz}

\begin{document}
\bibliographystyle{sysbio}
 \begin{abstract}
We analyse a maximum-likelihood approach for combining phylogenetic
trees into a larger `supertree'. This is based on a simple
exponential model of phylogenetic error, which ensures that ML
supertrees have a simple combinatorial description (as a median
tree, minimising a weighted sum of distances to the input trees). We
show that this approach to ML supertree reconstruction is
statistically consistent (it converges on the true species supertree
as more input trees are combined), in contrast to the widely-used
MRP method, which we show can be statistically inconsistent under
the exponential error model. We also show that this statistical
consistency extends to an ML approach for constructing species
supertrees from gene trees. In this setting, incomplete lineage
sorting (due to coalescence rates of homologous genes being lower
than speciation rates) has been shown to lead to gene trees that are
frequently different from species trees, and this can confound
efforts to  reconstruct the species phylogeny correctly.
 \end{abstract}

\maketitle

\bigskip
\section{Introduction}
Combining trees on different, overlapping sets of taxa into a parent
`supertree' is now a mainstream strategy for constructing large
phylogenetic trees. The literature on supertrees is growing steadily: new
methods of supertree reconstruction are being developed
\citep{cotton} and supertree analyses are shedding light on
fundamental evolutionary questions \citep{binindaemonds}.  Despite
this surge in research activity, it is probably fair to say that
biologists are still confused about what supertrees really are and
what it is we do when we build a supertree.  Are we, as some
maintain, simply summarising the phylogenetic information contained
in a group of subtrees? Or are we trying to derive the best estimate
of phylogeny given the information at hand?  Nor is it clear which
of these two conceptually different objectives underpin the various
supertree reconstruction methods.

We take the view that what biologists really want a supertree
reconstruction method to deliver is the best hypothesis of
evolutionary relationships that can be inferred from the data
available.  Obviously, it is not the case that the supertree
constructed as a summary statistic will necessarily be the best
estimate of phylogeny.   Nonetheless, if we are prepared to consider
supertree reconstruction a problem of phylogenetic
\textit{estimation}, we have at our disposal an arsenal of
phylogenetic tools and methods that have been tried and tested.
Matrix Representation with Parsimony (MRP; \citep{BaumRagan}),
Matrix Representation with Compatibility (MRC; \citep{rodrigo96,
RossRodrigo}) and, most recently, Bayesian supertree reconstruction
(BSR, \citep{Ronquistetal}) are undoubtedly inspired by standard
phylogenetic methods. A gap remains, though, as there has been
remarkably little development of likelihood-based methods for
supertree reconstruction.

In this paper, we analyse one approach to obtain maximum-likelihood
(ML) estimates of supertrees, based on a probability model that
permits `errors' in subtree topologies. The approach is of the type
described by Cotton and Page (2004), and it permits supertrees to be
estimated even if there is topological conflict amongst the
constituent subtrees. We show that ML estimates of supertrees so
obtained are statistically consistent under fairly general
conditions. By contrast, we show that MRP may be inconsistent under
these same conditions.  We then consider a further complication that
arises in the supertree setting when combining gene trees into
species trees - in addition to the possibility that the input gene
trees are reconstructed incorrectly (either a consequence of the
reconstruction method used, or some sampling error), there is a
further stochastic process that leads to the (true) gene trees
differing from their underlying species tree (a consequence of
incomplete lineage sorting).  Although simple majority-rule
approaches (and gene concatenation) have recently been shown to be
misleading, we show that an ML supertree approach for combining gene
trees is also statistically consistent.

\subsection{Terminology}

Throughout this paper, unless stated otherwise, phylogenetic trees
may be either rooted or unrooted, and we will mostly follow the
notation of Semple and Steel (2003). In particular, given a (rooted
or unrooted) phylogenetic tree $\T$ on a set $X$ of taxa (which will
always label the leaves of the tree), any subset $Y$ of $X$ induces
a phylogenetic tree on taxon set $Y$, denoted $\T|Y$, which,
informally, is the subtree of $\T$ that connects the taxa in $Y$
only. In the {\em supertree} problem, we have a sequence $\p =
(\T_1, \T_2, \ldots, \T_k)$ of input trees, called a {\em profile},
where $\T_i$ is a phylogenetic tree on taxon set $X_i$.  We wish to
combine these trees into a phylogenetic tree $\T$ on the union of
the taxon sets (i.e. $X= X_1 \cup X_2 \cup \cdots \cup X_k$). We
assume that the trees in $\p$ are either all rooted or all unrooted,
and that $\T$ is rooted or unrooted accordingly. We will mostly
assume that trees are {\em fully-resolved} (i.e. binary trees,
without polytomies); in Section~\ref{tech} we briefly describe how
this restriction can be lifted.

A special case of the supertree problem arises when the taxon sets
of the input trees are all the same ($X_1 = X_2 = \cdots =X_k$).
This is the much studied {\em consensus tree} problem.   In an early
paper McMorris (1990) described how, in this consensus setting, the
majority rule consensus tree can be given a maximum likelihood
interpretation.   However this approach is quite different to the
one described here (even when restricted to the consensus problem).

In this paper, we will denote the underlying (`true') species tree
as $\T_0$ (assuming that such a tree exists and that the evolution
of the taxa has not involved reticulate processes such as the
formation of hybrid taxa).  In an ideal world, we would like $\T_i =
\T_0|X_i$ for each tree $\T_i$ in the profile -- that is, we would
like each of the reconstructed trees to be identical to the subtree
of the `true' tree for the taxa in $X_0$. But in practice, the trees
$\T_1,...,\T_k$ are unlikely to even be compatible (i.e. no
phylogenetic tree $\T$ exists for which $\T_i = \T|X_i$ for all
$i$).

\section{An exponential model of phylogenetic error}

Species trees that have been inferred from data may differ from the
true underlying species tree for numerous reasons, including
sampling effects (short and/or site-saturated sequences, or poorly
defined characters), model violation, sequencing or alignment
errors, and so forth. In this section, we will assume a simple model
of phylogenetic error in which the probability of observing a given
tree falls off exponentially with its `distance' from an underlying
generating tree (e.g. the true species tree $\T_0$). This type of
model has been described by Holmes (2003), albeit from a different
perspective.   Suppose $d$ is some metric on resolved phylogenetic
trees.  In the exponential model, the probability, denoted
$\PP_{\T}[\T']$, of reconstructing any species tree $\T'$ on taxon
set $Y$, when $\T$ is the generating tree (on taxon set $X$) is
proportional to an exponentially decaying function of the distance
from $\T'$ to $\T|Y$. In other words,
\begin{equation}
\label{mallows} \PP_{\T}[\T'] = \alpha \exp(-\beta d(\T',\T|Y)).
\end{equation}
The constant $\beta$ can vary with $Y$ and other factors (such as
the quality of the data); for example, trees constructed from long
high-fidelity sequences are likely to have a larger $\beta$ than
trees constructed from shorter and/or noisier sequences.  The
constant $\alpha$ is simply a normalising constant to ensure that
$\sum_{\T'} \PP_{\T}[\T'] = 1$, where the sum is over all fully
resolved phylogenetic trees $\T'$ on taxon set $Y$. When we have a
sequence $(X_1, X_2, \ldots)$ of subsets of $X$, we will reflect the
dependence of $\alpha, \beta$ on $X_i$ by writing $\alpha_i,$ and
$\beta_i$. Note that $\alpha_i$ is determined entirely by $\beta_i$
and $|X_i|$.

Note that, implicit in (\ref{mallows}), the probability of $\T'$
depends only on the subtree of $\T$ connecting the species in $\T'$
and not on the other species in $\T$ that are not present in $\T'$.

Now, suppose we observe the profile of trees $\p=(\T_1, \T_2,
\ldots, \T_k)$ as above, where $\T_i$ has leaf set $X_i$.  Assume
that, for each $i$, the tree $\T_i$ has been independently sampled
from the exponential distribution (\ref{mallows}) with $\beta =
\beta_i$. Select a phylogenetic $X$--tree $\T$ that maximises the
probability of generating the observed profile $\p$ (we call this
type of a tree $\T$ an {\em ML supertree for $\p$}).  In the special
case where $d$ is the nearest-neighbor interchange (NNI) metric, and
the $\beta_i$ values are all equal, this ML supertree was described
by \citep{cot}.  An ML supertree has a simple combinatorial
description as a (weighted) median tree, as the following result
shows.

\begin{proposition}
\label{lem1} For any metric $d$ on phylogenetic trees, an ML
supertree for a profile $\p$ is precisely a tree $\T$ that minimises
the weighted sum:
$$\sum_{i=1}^k\beta_i d(\T_i, \T|X_i).$$
\end{proposition}
\begin{proof}
By the independence assumption
$$\PP_{\T}[(\T_1, \T_2, \ldots, \T_k)]= \prod_{i=1}^k\PP_{\T}[\T_i],$$ and, by
(\ref{mallows}), $\PP_{\T}[\T_i]=\alpha_i\exp(-\beta_i d(\T_i,
\T|X_i))$. Consequently, $\PP_{\T}[(\T_1, \T_2, \ldots, \T_k)]$ is
proportional to
$$\exp(-\sum_{i=1}^k \beta_id(\T_i, \T|X_i)),$$
and this is maximised for any tree $\T$ that minimises $\sum_{i=1}^k
\beta_id(\T_i, \T|X_i).$ This completes the proof.
\end{proof}

Notice that in the consensus tree setting, and where $d$ is the
symmetric difference (Robinson-Foulds) metric, the consensus of the
ML supertrees is the same as the usual majority rule consensus tree.
This follows from earlier results by Bath{\'e}lemy and McMorris (1986) \citep[see also][]{cotton}.

\section{Statistical Consistency of ML supertrees under the exponential model}

Is the ML procedure statistically consistent as the number ($k$) of
trees in the profile grows? More precisely, under what conditions is
the method guaranteed to converge on the underlying generating tree
$\T_0$ as we add more trees to the analysis? The problem is slightly
different from other settings (such as the consistency of ML for
tree reconstruction from aligned sequence data) where one has a
sequence of i.i.d. random variables. In the supertree setting, it is
perhaps unrealistic to expect that the data-sets are generated
according to an identical process, since the sequence of subsets
$X_1, X_2,\ldots $ of $X$ is generally deliberately selected.

To formalise the statistical consistency question in this setting,
let $X_1, X_2,\ldots, $ be a sequence of subsets of $X$. It is clear
that the $X_i$'s must
 cover $X$ in some `reasonable' way in order for the ML supertree method to be consistent
  -- for example, if some taxon is not present in any
$X_i$, or is present in only a small number of input trees, then we
cannot expect the location of this taxon in any supertree to be
strongly supported.

Thus, we will assume that the sequence of subsets of $X$ satisfies
the following {\em covering property}: For each subset $Y$ of taxa
from $X$ of size $m$ (where $m =3$ for rooted trees or $m=4$ for
unrooted trees), the proportion of subsets $X_i$ that contain $Y$
has strictly positive support as the sequence length (of subsets)
increases. More formally, for each such subset $Y$ of $X$ we assume
there is some $\epsilon  >0$ and some $K$ sufficiently large for
which:
\begin{equation} \label{cover}
\frac{1}{k} |\{i\leq k: Y \subseteq
X_i\} |\geq \epsilon \mbox{ for all $k \geq K.$}
\end{equation}
If a subset of taxa, $Y$, is only found in one or a few trees, and
is never seen again in trees that are subsequently added, this
property will not hold.

Now, suppose we sample a random tree $\T_i$ on leaf set $X_i$
according to the exponential distribution (\ref{mallows}). Let $\p_k
=(\T_1,...,\T_k)$ be the resulting profile of independently sampled
trees. The following theorem establishes the statistical consistency
of ML supertrees under the exponential model, when the covering
condition property holds.

\begin{theorem}
\label{mlcons} Given a sequence $X_1,X_2,\ldots$ which satisfies the
covering property (\ref{cover}), consider a profile $\p_k= (\T_1,
\ldots, \T_k)$, where $\T_i$ is generated independently according to
the exponential model (\ref{mallows}) with $\beta = \beta_i$ and
with generating tree $\T_0$.  Suppose that $\beta_i \geq \delta>0$
for all $i$. Then the probability that $\p_k$ has a unique ML
supertree and that this tree is $\T_0$ tends to $1$ as $k
\rightarrow \infty$.
\end{theorem}
\begin{proof}
To establish the theorem, using Proposition~\ref{mlsup} (stated and
proved in the Appendix) it is enough to specify for each choice of
distinct resolved phylogenetic $X$--trees $\T_0$ and $\T$, a
sequence of events $E_k$ (dependent on $\p_k$) for which, as $k$
grows, $E_k$ has a probability that tends to $1$ under the
distribution obtained from $\T_0$ and tends to $0$ under the
distribution obtained from $\T$. Since $\T$ differs from $\T_0$ a
subset $Y$ exists of size $m$ ($=3$ for rooted trees and $=4$ for
unrooted) for which $\T|Y \neq \T_0|Y$. Notice that the covering
property (\ref{cover}) implies that
\begin{equation} \label{cover2}
\frac{1}{k} |\{i\leq k: \T|X_i \neq \T_0|X_i\} |\geq \epsilon  \mbox{ for all $k \geq K.$}
\end{equation}
Let $E_k$ be the event that among all those $i \in \{1, \ldots, k\}$
for which $\T|X_i \neq \T_0|X_i$ we have $\T_i = \T_0|X_i$ more
often than $\T_i = \T|X_i$. Now, for a profile generated by $\T_0$
according to the exponential model (\ref{mallows}), we have
$$\PP_{\T_0}[\T_i= (\T_0|X_i)]=\alpha_i\exp(-\beta_i \cdot 0) = \alpha_i,$$
and for each $i$ for which  $\T|X_i \neq \T_0|X_i$, we also have
$$\PP_{\T_0}[\T_i= (\T|X_i)] \leq \alpha_i\exp(-\delta d(\T|X_i, \T_0|X_i)).$$
In particular, for each $i$ for which $\T|X_i \neq \T_0|X_i$,
\begin{equation}
\label{pbound1}
\PP_{\T_0}[\T_i= (\T_0|X_i)] \geq (1+\eta)\PP_{\T_0}[\T_i= (\T|X_i)] \mbox {  for some $\eta>0$.}
\end{equation}
Similarly, for a profile generated by $\T$ according to the
exponential model (\ref{mallows}) and for each $i$ for which $\T|X_i
\neq \T_0|X_i$, we have
\begin{equation}
\label{pbound2}
\PP_{\T}[\T_i= (\T|X_i)] \geq (1+\eta)\PP_{\T}[\T_i= (\T_0|X_i)]  \mbox{ for some $\eta>0$.}
\end{equation}

By condition (\ref{cover2}), there is a positive limiting proportion
($\epsilon>0$) of $i$ for which $\T|X_i \neq \T_0|X_i$; therefore
inequalities (\ref{pbound1}) and (\ref{pbound2})
 imply that event $E_k$ has a probability tending to
$1$ (or $0$) as $k \rightarrow \infty$ for a profile generated by
$\T_0$ (or $\T$ respectively) as required. Statistical consistency
of ML now follows by Proposition~\ref{mlsup}.
\end{proof}

\section{Relation to MRP and its statistical inconsistency}
As shown recently by Bruen and Bryant (2007), there is a close
analogy between MRP (Matrix Representation with Parsimony) and
consensus tree methods which seek a median tree computed using the
SPR (subtree prune and regraft)  or TBR (tree bisection and
reconnection) metric $d$ (recall that a median tree for a profile
$\p = (\T_1, \ldots \T_k)$ of trees that all have same leaf set $X$,
is a tree $\T$ that minimises the sum $\sum_{i=1}^k d(\T, \T_i)$;
{\em cf.} Proposition~\ref{lem1}). However, the result from Bruen
and Bryant (2007) does not guarantee that MRP produces an ML
supertree even when $\beta_i = 1$ for all $i$.

We turn now to the question of the statistical consistency of MRP
under the exponential model (\ref{mallows}). It can be shown that
MRP will be statistically consistent under the covering property
(\ref{cover}) in some special cases. Two such cases that can be
formally established (details omitted) are: (i) when all the subsets
$X_i$ are of size $4$, or (ii) when $\beta_i$ is sufficiently large
(in relation to $|X|$). However, in general, we have the following
result.

\begin{theorem}
\label{incon} A $\beta>0$ exists for which MRP is statistically
inconsistent even in the special (`consensus') case where, for all
$i$, $X_i$ is the same set of six taxa and $\beta_i = \beta$. More
precisely, for this value of $\beta$ and with unrooted
fully-resolved phylogenetic trees on these (equal) taxon sets, the
probability that $\T_0$ is an MRP tree (for a profile of trees
generated under (\ref{mallows})) converges to $0$ as $k$ tends to
infinity.
\end{theorem}

\begin{proof}
For two unrooted fully-resolved phylogenetic $X$--trees $\T, \T'$
let $L(\T,\T')$ denote the total parsimony score on $\T'$ of the
sequence of splits of $\T$.  That is,
\begin{equation}
\label{equatione1}
L(\T,\T') = \sum_{\sigma \in \Sigma(\T)} l(\sigma, \T'),
\end{equation}
where $\Sigma(\T)$ is the set of splits of $\T$ and $l(\sigma, \T')$
is the parsimony score of the split $\sigma$ on $\T'$ (treating
$\sigma$ as a binary character, \citep{sem}). For any fully-resolved
phylogenetic $X$--tree $\T'$ let $e(\T'; \T_0)$ be the expected
total parsimony score on $\T'$ of the sequence of splits of a tree
$\T$ randomly generated by $\T_0$ according to the exponential model
(\ref{mallows}). Then,
\begin{equation}
\label{equatione2}
e(\T';\T_0)= \sum_{\T} \alpha \exp(-\beta d(\T, \T_0))\cdot L(\T,\T').
\end{equation}
To establish Theorem~\ref{incon}, it is enough to show, for some
$\beta>0$ and for two unrooted fully-resolved trees $\T_0, \T_1$ on
$X= \{1, \ldots, 6\}$, that $e(\T_0; \T_0) - e(\T_1; \T_0) > 0,$
since if $\T_0$ is the generating tree, then $\T_1$ will be favored
over $\T_0$ by MRP. We first show that this can occur when $\beta =
0$. In that case, $\alpha\exp(-\beta d(\T, \T_0))=1/105$ for all
$\T$ (there are $105$ unrooted fully-resolved phylogenetic trees on
$X$) and so, by (\ref{equatione1}) and (\ref{equatione2}),
$$e(\T_0; \T_0) - e(\T_1; \T_0) = \frac{1}{105}\sum _{\T}(L(\T, \T_0) - L(\T, \T_1)) = \frac{1}{105}\sum_{\sigma}n(\sigma)\cdot (l(\sigma, \T_0)- l(\sigma, \T_1)),$$
where $n(\sigma)$ is the number of unrooted fully-resolved
phylogenetic $X$--trees containing split $\sigma$ and the summation
is over all the splits of $X = \{1,\ldots, 6\}$. Now suppose $\T_0$
has a symmetric  shape (i.e. an unrooted fully-resolved tree of six
leaves with three cherries) and $\T_1$ has a pectinate shape (i.e.
an unrooted fully-resolved tree of six leaves with two cherries).
Then, by using earlier results \citep[Table 3]{sig} and basic
counting arguments, it can be shown that
$$e(\T_0; \T_0) - e(\T_1; \T_0) = \frac{1}{21}.$$
So far, we have assumed that $\beta =0$, however, $e(\T_0; \T_0) -
e(\T_1; \T_0)$ is a continuous function of $\beta$ so a strictly
positive value of $\beta$ exists for which $$e(\T_0; \T_0) - e(\T_1;
\T_0) \geq \frac{1}{10}.$$ This  completes the proof.
\end{proof}

\noindent An interesting theoretical question is whether a value $s
\in (0,1)$ exists for which MRP is statistically consistent (for
arbitrarily large taxon sets) under the conditions of Theorem 4.1,
whenever $\beta \geq s$.

\section{Technical remarks}
\label{tech}

{\em Extension to trees with polytomies.}

\noindent We can easily modify the ML process if some of the input
trees are not fully-resolved.  For a general phylogenetic tree $t_i$
(possibly with polytomies) on taxon set $X_i \subseteq X$, and a
generating fully-resolved phylogenetic tree $\T$ on taxon set $X$,
let $\phi(t_i|\T)$ be the probability of the event that the tree
$\T_i$ that $\T$ generates under the exponential model is a
refinement of $t_i$. More precisely,
$$\phi(t_i|\T) = \sum_{\T_i \geq t_i} \alpha_i\exp(-\beta_i d(\T_i, \T|X'))$$
where $\T_i \geq t_i$ indicates that the (fully-resolved) tree
$\T_i$ contains all the splits present in $t_i$, and has the same
leaf set ($X_i$). Notice that $\phi(t_i|\T)$ is not a probability
distribution on phylogenetic trees with the leaf set $X'$ (its sum
is $>1$). Nevertheless, given a profile $\p= (t_1, \ldots, t_k)$ of
phylogenetic trees (some or all of which may have polytomies), we
can perform ML to select the tree $\T$ that maximises the joint
probability  $\prod_{i=1}^k \phi(t_i|\T)$ of the events $\T_i \geq
t_i$ for $i =1,\ldots, k$.

{\em An alternative perspective on ML supertrees for certain tree
metrics.}

\noindent We point out an alternative way of viewing this ML
procedure applied to a profile $\p = (\T_1, \ldots, \T_k$) when $d$
is one of two well-known metrics on trees (SPR and TBR). Suppose
that we were to extend each tree $\T_i$ in $\p$ to a tree $\T'_i$ on
the full set of taxa ($X$). We could regard the placement of those
taxa that are `missing' in $\T_i$ (namely the taxa in $X-X_i$) to
form a tree $\T_i'$ on the full leaf set $X$ to be `nuisance
parameters' in a maximum likelihood framework (under the exponential
model), and thereby seek to find the tree $\T$ and extensions
$(\T_1', \ldots, \T_k')$ to maximise the joint probability:
$$\PP_{\T}[(\T_1', \T_2', \ldots, \T_k')] \mbox{ subject to } \T_i = \T_i'|X_i
\mbox{ for all $i$}.$$

We call such a tree $\T$ an {\em extended ML tree} for the profile $\p$.

\begin{proposition}
For $d = SPR$ or $d=TBR$, and any profile $\p$ of fully-resolved,
unrooted phylogenetic trees, the extended ML tree(s) for $\p$
coincides precisely with the ML tree(s) for $\p$.
\end{proposition}
\begin{proof}
For $d = SPR$ or $d=TBR$, it can be shown that for any resolved
unrooted phylogenetic trees $\T$ on leaf set $X$, and $\T_i$ on leaf
set $X_i$, that:
\begin{equation}
\label{identity} \min\{d(\T_i', \T): \T_i'|X_i = \T_i\} = d(\T_i,
\T|X_i).
\end{equation}
The result now follows by Proposition~\ref{lem1}
\end{proof}
Note that Equation~\ref{identity} does not necessarily hold for
other tree metrics (such as the NNI (nearest-neighbor interchange)
or the partition (Robinson-Foulds) metric).

\section{Statistical consistency of ML species supertrees from multiple gene trees}

A current problem in phylogenetics is how best to infer species
trees from gene trees \citep{deg, gad, liu}. Even in the consensus
setting (i.e. when the set of taxa for each gene tree is the
complete set of taxa under study), Degnan and Rosenberg (2006)
have demonstrated how incomplete lineage sorting on gene trees can
mean that the most likely topology for a gene tree can differ from
the underlying species tree (for any certain rooted phylogenetic
trees on four taxa and for all rooted phylogenetic trees on five or
more taxa). This surprising result implies that simplistic `majority
rule' approaches to finding a consensus species tree can be
problematic.

The phenomenon described by Degnan and Rosenberg (2006) is based on the coalescent
model for studying lineage sorting in evolving populations. The
surprising behavior arises only when the effective population sizes
and the branch lengths of the species tree are in appropriate
ranges. Moreover, for $3$--taxon trees, the most probable gene tree
topology always agrees with the species tree topology. Nevertheless,
the fact that larger gene trees can favour an incorrect species tree
might easily complicate some standard statistical approaches.

In this section, we show how, despite the phenomena described above
\citep[from][]{deg}, and even in the more general supertree setting
(where some gene trees may have some missing taxa), a maximum
likelihood approach to supertree construction of a species tree from
gene trees is statistically consistent. Moreover, we frame this
approach so that it is sufficiently general to also allow for error
in the reconstruction of the gene trees (as arises under the
exponential model).

Consider a model $M$ that has a generating tree topology $\T$ as its
sole underlying parameter. Such a model will typically derive from a
more complex model containing other parameters (such as branch
lengths, population sizes and so forth), but we will assume that
these have a prior distribution and that they have been integrated
out, so our model has just one parameter - the tree topology. We say
that $M$ satisfies the property of {\em basic centrality} if, for
all subsets of $Y$ of $X$ of size $m$ ($=3$ for rooted trees and
$=4$ for unrooted trees), we have:
\begin{equation}
\label{lineage} \PP_{\T}[\T|Y] \geq (1+ \eta)\PP_{\T}[\T']
\end{equation} for all trees $\T'$ on leaf set $Y$ that are different from $\T|Y$, and where $\eta>0$.
For example, the exponential model (\ref{mallows}) satisfies basic
centrality, since (\ref{lineage}) holds for {\em all} subsets $Y$ of
$X$.  For lineage sorting (with a prior distribution on ancestral
population sizes and branch lengths), the property holds, but only
because (\ref{lineage}) holds for the subsets $Y$ of $X$ of size
$3$, as we describe shortly. Firstly, however, we state a
statistical consistency result that extends Theorem~\ref{mlcons}.

\begin{proposition}
Given a sequence $X_1,X_2,\ldots$ which satisfies the covering
property (\ref{cover}), consider a profile $\p_k= (\T_1, \ldots,
\T_k)$, where $\T_i$ is generated independently according to a model
that satisfies the basic centrality property (with $\T=\T_i$, and
$\eta_i
> \delta>0$ for all $i$) and with generating tree $\T_0$. Then the probability that $\p_k$ has a
unique ML supertree and that this tree is $\T_0$ tends to $1$ as $k
\rightarrow \infty$.
\end{proposition}
\begin{proof}
The proof is similar to the proof of Theorem~\ref{mlcons}, the
essential difference being a modification to the way the events
$E_k$ are defined.  Given $\T_0$ and $\T$ (as in the proof of
Theorem~\ref{mlcons}), let $E_k$ be the event that for each $i \in
\{1, \ldots, k\}$ for which $Y \subseteq X_i$ we have $\T_i|Y =
\T_0|Y$ more often than $\T_i|Y = \T|Y$.  Then (as in the proof of
Theorem ~\ref{mlcons}) as $k$ grows, $E_k$ has a probability that
converges to $1$ if $\T_0$ is the generating tree, and a probability
that converges to $0$ if $\T$ is the generating tree. The theorem
now follows by Proposition~\ref{mlsup}.
\end{proof}

We now apply this result in a supertree setting where we have the
compounding effect of {\em two} sources of error: (i) error in using
the true gene tree to represent the true species tree, due to
lineage sorting, and (ii) error in reconstructing the correct gene
tree, modeled by the exponential model (\ref{mallows}).  We claim
that an ML approach to inferring a species tree in the presence of
these two sources of error is still statistically consistent (under
the coalescent and exponential model (\ref{mallows}), and assuming the covering
property), due to the following argument, which justifies the basic
centrality property.

Consider a rooted fully-resolved species tree $\T$ on $X$ and a
rooted full-resolved gene tree $\T'$ on $Y$, where $Y$ is a subset
of $X$ of size 3 (note that we are here identifying the taxa in the
gene tree with taxa in the species tree). Then the probability of
observing $\T'$ under the combination of these two sources of error
(treated independently) from a generating species tree $\T$ can be
written as
$$\PP_{\T}[\T'] = \sum_{\T''}\PP^c_{\T}[\T'']\PP_{\T''}[\T'] ,$$
where the summation is over the three rooted fully-resolved gene
trees on taxon set $Y$, $\PP^c_{\T}[\T'']$ is the probability that
species tree $\T$ gives rise to gene tree $\T''$ for the taxa on $Y$
(under lineage sorting according to the coalescent model), and
$\PP_{\T''}[\T']$ is the probability that a generating gene tree
$\T''$ produces $\T'$, as given by the exponential model
(\ref{mallows}). Now, considering lineage sorting under the
coalescent model, we have $\PP^c_{\T}[\T|Y]=\frac{1}{3}(1 + 2\tau)$
for $\tau \in (0,1)$ while $\PP^c_{\T}[\T'']= \frac{1}{3}(1 -\tau)$
for the two other choices of $\T'' \neq \T|Y$  \citep[see e.g.,][]{ros,
taj}. Furthermore, under the exponential model (\ref{mallows}), and
assuming, without loss of generality, that $d$ takes the value $0$
or $1$ for each pair of 3-taxon trees, we have $\PP_{\T''}[\T''] =
\alpha$, while $\PP_{\T''}[\T'] = \alpha e^{-\beta}$ for the other
two choices of $\T' \neq \T''$ (and $\alpha =
(1+2e^{-\beta})^{-1}$).  Combining these relationships, we obtain:
$$\PP_{\T}[\T|Y] =\frac{\alpha}{3}\left(1 + 2\tau+ 2(1-\tau)e^{-\beta}\right),$$
while for the other two choices of $\T' \neq \T|Y$, we have
$$\PP_{\T}[\T'] =\frac{\alpha}{3}\left((1 + 2\tau)e^{-\beta} +
(1-\tau)(1+e^{-\beta})\right).$$ For any given $\beta, \tau
>0$, these last two equations imply that for some $\eta>0$, and for
the two choices of $\T' \neq T|Y$, we have
$$\PP_{\T}[\T|Y] \geq (1+ \eta)\PP_{\T}[\T']$$
(in fact, routine algebra shows that we can take $\eta =
3\tau(1-e^{-\beta})$). Taking the value of $\eta$ that is minimal
for all subsets $Y$ of $X$ of size $3$ establishes the basic
centrality property in this setting.

\section{Discussion}

To develop a likelihood-based supertree reconstruction method, it is
necessary to define a model that delivers the probability of
obtaining a series of subtree topologies, given a hypothesised
supertree.  We have chosen a very simple yet intuitive probability
function whereby the probability of observing a `wrong' subtree
(i.e. one where the topology differs from that of a pruned
supertree) decreases exponentially as its topology becomes
increasingly distant from that of the hypothesised supertree.
Consequently, the ML supertree can be estimated even when the
constituent subtrees have conflicting topological signals.

Our approach is model-based, but one may reasonably ask whether the
model described here is a biologically realistic one. We suggest that it is.
  For one thing, we expect, for a variety of reasons, to
see conflicts between the topologies of subtrees and the
reconstructed supertree.  With gene sequences obtained from
different species, for instance, incomplete lineage sorting and
ancestral heterozygosity frequently lead to differences between gene
trees and species trees. Convergent and parallel evolution can
confound phylogenetic reconstruction, as can long-branch attraction.
We have chosen to use the exponential distribution to describe this
steady decrease in probabilities as the distances between subtrees
and supertrees increase.  The value of using the exponential
distribution lies in the ease with which it can be manipulated when
we compute log-likelihoods.  However, we suggest that one fruitful
research project may be to explore other possible probability
distributions and, for that matter, other tree-to-tree distance
metrics.

The likelihood framework provides an additional benefit: a rich body
of statistical and phylogenetic methods already use likelihood.
Moreover, statistical consistency holds for maximum likelihood
supertrees under weak conditions, in contrast to MRP, which can be
inconsistent in some cases.  We also show that the ML supertree
approach developed here provides a statistically consistent strategy
for combining gene trees even when there is the possibility that
these trees may be different from the true species tree.  An obvious
application of ML supertrees will be their use in statistical tests
of topological hypotheses, and we already know how to do this with
standard ML phylogenies \citep{gold}.

We also recognise that our particular likelihood implementation is
closely related to the `Majority-Rule(-) Supertree' construction
proposed by Cotton and Wilkinson (2007)). More precisely,
when the tree metric is the symmetric difference (Robinson-Foulds)
metric, then the Marjority-Rule(-) Supertree is, in effect, the
strict consensus of our ML supertrees. However, the approach in
Cotton and Wilkinson (2007) is quite different: they show how to extend
majority-rule from the consensus to the supertree setting.
Nonetheless, they converge on the same optimality criterion that we
use, i.e. a supertree that minimises the sum of distances to a set
of trees. One should not be surprised that the same optimality
criterion can emerge from different conceptual bases.  With standard
phylogenetic reconstruction, choosing the tree that minimises the
number of evolutionary changes can be justified philosophically
(with the principle of maximum parsimony) as a consensus method
\citep{bru}, or by using an explicitly statistical approach (e.g.
likelihood \citep{steel}).

We have not discussed algorithms to search for ML supertrees.  Instead, we direct readers to the discussion in Cotton and Wilkinson (2007), since the criterion we use is similar to theirs.

\section{Appendix:  Consistency of ML for general (non-i.i.d.) sequences}

Here we describe a convenient way to establish the statistical
consistency of maximum likelihood when we have a sequence of
observations that may not be independent or identically distributed.
We frame this discussion generally, as the result may be useful for
other problems. In particular, in this result, we do not need to
assume the sequence samples are independent (though in our
applications, they are), nor identically distributed (in our
applications, they are not). Suppose we have a sequence of random
variables $Y_1, Y_2,...$ that are generated by some process that
depends on an underlying discrete parameter $a$ that can take values
in some finite set $A$. In our setting, the $Y_i$'s are trees
constructed from different data sets (e.g. gene trees), while $a$ is
the generating species tree topology. We assume that the model
specifies the probability distribution of $(Y_1,\ldots, Y_k)$ given
(just) $a$ -- for example, in our tree setting this would mean
specifying prior distributions on the branch lengths and other
parameters of interest (eg. ancestral population sizes) and
integrating with respect to these priors.

Given an actual sequence $(y_1,..., y_k)$ of observations, the {\em
maximimum likelihood (ML) estimate} of the discrete parameter is the
value $a$ that maximises the joint probability $$\PP_a[Y_1 = y_1,
\ldots, Y_k = y_k]$$  (i.e. the probability that the process with
parameter $a$ generates $(y_1, \ldots, y_k)$). Now suppose that the
sequence $Y_1,\ldots Y_k, \ldots $ is generated by $a_0$. We would
like the probability that the ML estimate is equal to $a_0$ to
converge to $1$ as $k$ increases.  If this holds for all choice of
$a_0 \in A$, then ML is {\em statistically consistent}. The
following result provides an convenient way to establish this;
indeed, it characterises the statistical consistency of ML.

\begin{proposition}
\label{mlsup}
 In the general set-up described above, ML is statistically consistent if and only if the following condition holds:
 for any two
distinct elements $a,b\in A$, we can construct a sequence of events
$E_1, E_2,\ldots$, where $E_k$ is dependent on $(Y_1, \ldots Y_k$),
for which, as $k \rightarrow \infty$:
\begin{itemize}
\item[(i)] the probability of $E_k$ under the model with
parameter $a$ converges to $1$.
\item[(ii)] the probability of $E_k$ under the model with
parameter $b$ converges to $0$.
\end{itemize}
\end{proposition}

\begin{proof}
The `only if' direction is easy: Suppose ML is statistically
consistent and $a, b \in A$ are distinct.  Let $E_k$ be the event
that $a$ is the unique maximum likelihood estimate obtained from
$(Y_1, \ldots, Y_k)$.  Then $E_k$ satisfies conditions (i) and (ii).

For the converse direction, recall that the {\em variation distance}
between two probability distributions $p,q$ on a finite set $W$ is
$$\max_{E \subset W} |\PP_p(E)-\PP_q(E)|$$ where $\PP_p(E) = \sum_{w
\in E} p(w)$ is the probability of event $E$ under distribution $p$
(similarly for $\PP_q(E))$.  This variation distance can also be
written as $\frac{1}{2}\|p-q\|_1$, where $\|p-q\|_1 =\sum_{w \in
W}|p(w) - q(w)|$ is the $l_1$ distance between $p$ and $q$.  Thus,
if we let $d^{(k)}(a,b)$ denote the $l_1$ distance between the
probability distribution on $(Y_1, \ldots, Y_k)$ induced by $a$ and
by $b$, then conditions (i) and (ii) imply that
\begin{equation}
\label{limit} \lim_{k \rightarrow \infty}\frac{1}{2}d^{(k)}(a,b)=1.
\end{equation}
Now, by the first part (Eqn. 3.1) of Theorem 3.2 of \citep{inv}, the
probability that the ML estimate is the value of $A$ that generates
the sequence $(Y_1, \ldots, Y_k)$ is at least $1 - \sum_{b \neq a}
(1- \frac{1}{2}d^{(k)}(a,b))$ and so, by (\ref{limit}), this
probability converges to $1$ as $k \rightarrow \infty.$
\end{proof}

\bibliography{Steel_Rodrigo}

\end{document}